\documentstyle[11pt,epsf]{article}

\newcommand{\ignorar}[1]{}

\newcommand{\ba}{\begin{eqnarray}}
\newcommand{\ea}{\end{eqnarray}}
\newcommand{\be}{\begin{equation}}
\newcommand{\ee}{\end{equation}}

\newcommand{\ga}{\mbox{$\!\!\!\!$}}
\newcommand{\gaga}{\mbox{\ga\ga\ga\ga}}

\newcommand{\alfa}{\mbox{\large$\bf \alpha $}}
\newcommand{\ai}{a}
\newcommand{\bi}{b}

\newcommand{\xiv}{\mbox{$\vec{\xi}\:$}}
\newcommand{\vv}{\mbox{$\vec{v}$}}
\newcommand{\hv}{\mbox{$\vec{h}$}}
\newcommand{\hvu}{\mbox{$\vec{h}^{\, 0}$}}
\newcommand{\hvd}{\mbox{$\vec{h}^{\, 1}$}}
\newcommand{\uv}{\mbox{$\vec{u}$}}
\newcommand{\wv}{\mbox{$\vec{w}$}}
\newcommand{\modd}[1]{\mbox{$ | #1  | ^{2}$}}

\newcommand{\Jm}{\mbox{\bf J}}
\newcommand{\Km}{\mbox{\bf K}}
\newcommand{\Lm}{\mbox{\bf L}}
\newcommand{\Gm}{\mbox{\bf C}}
\newcommand{\Gammam}{\mbox{$\bf \Gamma $}}
\newcommand{\Mm}{\mbox{\bf M}}
\newcommand{\Deltam}{\mbox{$\bf \Delta$}}
\newcommand{\Um}{\mbox{\bf U}}
\newcommand{\Utm}{\mbox{$\bf \tilde{U} $}}
\newcommand{\Vtm}{\mbox{$\bf \tilde{V} $}}
\newcommand{\bM}{\mbox{$\bar{M}$}}
\newcommand{\sbM}{\mbox{\scriptsize $\bar{M}$}}
\newcommand{\Ggm}{\mbox{$\bf\cal G $}}
\newcommand{\id}[1]{\mbox{${\bf Id\/}_{#1 \times #1}$}} 
\newcommand{\unom}{\mbox{\bf\Large $1\!\! 1$ } }

\newcommand{\prob}{\mbox{\Large $\wp $}}
\newcommand{\bitent}[1]{\mbox{${\cal H\/} ( #1 ) $}}
\newcommand{\hbitent}[1]{\mbox{${\cal {H}^{^{^{\ga\wedge\;}}} \/}(#1)$}}
\newcommand{\hr}[1]{\mbox{$\hat{\rho }(#1)$}}
\newcommand{\ta}[1]{\mbox{$\prec\!\!\prec #1 \succ\!\!\succ $}}

\newcommand{\erf}[1]{\mbox{$ er\! f\! (#1) $}}

\newcommand{\Integer}{\mbox{$\bf \cal{Z} $}}
\newcommand{\Real}{\mbox{$\Re $}}

\newcommand{\matreal}[2]{\mbox{${\it M \/} _{ #1 \times #2 } (\Real )$}}
\newcommand{\sint}[2]{\mbox{$\int d^{ #1 } #2 \;\; $}}
\newcommand{\sapprox}[1]{\mbox{$_{ _{ _{_{#1}}}}
\ga\ga\!\!\approx\, $}}
\newcommand{\lapprox}[1]{\mbox{$ _{_{_{_{ #1}}}}
\gaga\approx\;\;\; $}}

\newcommand{\ii}{\mbox{\large\it i}}

\title{The mutual information of a stochastic binary 
channel: validity of the Replica Symmetry Ansatz}

\author{{\bf\sc Antonio Turiel}, {\bf\sc 
Elka Korutcheva\thanks{Permanent address: 
G.Nadjakov Institute of Solid State Physics,
Bulgarian Academy of Sciences, 1784 Sofia, Bulgaria. 
\newline
\indent
Present address: 
Departamento de F\'{\i}sica Fundamental,
Universidad Nacional de Educaci\'on a Distancia,
c/ Senda del Rey s/n,
28080 Madrid, Spain
}}  
and {\bf\sc N\'estor Parga}\\
\it Departamento de F\'{\i}sica Te\'orica \\
\it Universidad Aut\'onoma de Madrid \\
\it Canto Blanco, 28049 Madrid, Spain }

\date{}

\begin{document}

\maketitle       

\begin{abstract}

We calculate the mutual information (MI) of a two-layered neural
network with noiseless, continuous inputs and binary, stochastic
outputs under several assumptions on the synaptic efficiencies.  The
interesting regime corresponds to the limit where the number of both
input and output units is large but their ratio is kept fixed at a
value $\alfa$. We first present a solution for the MI using the
replica technique with a replica symmetric (RS) ansatz.  
Then we find an exact solution for this quantity valid 
in a neighborhood of $\alfa = 0$.
An analysis of this solution shows that the
system must have a phase transition at some finite value of
$\alfa$. This transition shows a singularity in the third derivative
of the MI.  As the RS solution turns out to be infinitely
differentiable, it could be regarded as a smooth approximation to the
MI. This is checked numerically in the validity domain of the exact
solution.

\end{abstract}                                    

P.A.C.S. 05.20; 87.30

Short title: Mutual Information of a Stochastic Network
                                                          
To appear in {\it Journal of Physics A}

\pagebreak

\section{Introduction}

\indent

The aim of this work is to study the properties of a binary
communication channel processing data from a Gaussian source, when the
output state is stochastic.  The architecture is a two-layered
feedforward neural network with $N$ analogue input units and $P$
binary output units. The mutual information (MI) is evaluated in the
large $N$ limit with $\alfa = \frac{P}{N}$ fixed.  
Research in this direction was previously done in
\cite{NP1}, where the case of a noiseless binary channel was studied,
and in \cite{KNP} which dealt with the case of a Gaussian source
corrupted with input noise.

The main motivation of this work is a technical one. In \cite{NP1} 
and \cite{KNP} the MI of binary channels was
obtained by means of the replica technique and the replica symmetry
ansatz \cite{Mezard}. However there have not been attempts to show the
validity of this solution.  In this paper we give an analytical
solution of the MI of the channel without making use
of the replica technique. In order to compare both methods, the replica
symmetry ansatz (RSA) solution of a general stochastic binary channel
is also evaluated.  While the RSA yields an expression of the MI for
all values of $\alfa$, the exact analytical solution turns out to be
valid only up to some $\alfa = O(1)$. However, our conclusion is that
the correct solution is the analytical one and that there is a
(possibly large order) phase transition located at the value of
$\alfa$ where the analytical solution ceases to be valid.  The RSA
solution has to be regarded just as a smooth approximation to the MI,
interpolating between the correct small and large $\alfa$ regimes.

There are several other motivations for doing this investigation. Once the MI 
of the channel is known, the 
problem of extracting as much information as possible from
the inputs can be addressed. This optimization problem 
leads to interesting data analysis.
Optimizing the MI, a criterion known as the ``infomax''
principle \cite{linsker88}, is a way of unsupervised learning (see,
e.g., \cite{Hertz}). The parameters of the model (that is, of the
channel) adapt according to this principle and in this way they learn
the statistics of the environment (that is, of the source).  
One can also optimize the MI by adapting the transfer function 
itself \cite{laughlin,low_noise}.
Another form of this type of unsupervised learning is the minimum
redundancy criterion \cite{Barlow}. Both have been used to predict the
receptive fields of the early visual system 
\cite{atickrevue,atick92,hateren,hateren2}.
The relation between them has been discussed
in ref.\cite{low_noise}. Another motivation is that learning how to 
solve this particular non-linear channel could provide the techniques to 
deal with other type of non-linearities. Little is known
 on the properties of systems other than linear,  except for
threshold-linear networks \cite{Treves95,SPRT97,Schultz98}
(treated with the replica technique),
approximations for weak non-linear terms in the processing
\cite{linsker93}, some general properties of the low and large
noise limits \cite{low_noise,Schuster}
and an analytical treatment of binary commutication channels 
either noiseless \cite{NP1} or with an input noise \cite{KNP}.

\indent
The paper is organized in the following way: The model is explained in
the next Section, where the notation and the relevant quantities are
also given.  In Section~\ref{I2_presentation}, the exact calculation of one 
of the contributions to the MI (the ``equivocation'' term, see 
\cite{Shannon}) is 
presented.  In Section~\ref{I1} the evaluation of the other contribution 
to MI (the entropy term) is discussed. In its first subsection, this is 
done by the usual replica technique.  The exact solution is obtained in the
second subsection.  In Section~\ref{expansions} the RS expression of
the MI is analysed in several asymptotic regimes. A numerical analysis
of the RSA solution is also presented at the end of this section.  The
comparison between the exact and the RSA solutions is done in
Section~\ref{comparison}. A discussion about the existence of a phase
transition is given in Section~\ref{phase}. Section~\ref{SB} is
devoted to the analysis of the Replica Symmetry Breaking (RSB)
solutions. The conclusions are contained in Section~\ref{conclusions}.
Finally several technical questions are presented in the Appendices.

\section{The model}
\label{model}

\indent

We consider a two-layered neural network with N inputs $\xiv\in\Real^{N}$
and P binary outputs $\vv\in (\Integer_{2})^{P}$. 
The input vector $\xiv $ is distributed as a Gaussian with zero mean  and 
covariance matrix $\Gm\:\in\:\matreal{N}{N}$:

\be \rho(\xiv )= \frac{e^{-\frac{1}{2} \xiv (\Gm )^{-1} \xiv^t }}
{\sqrt{\det{(2 \pi \Gm)}}}.
\label{xivdistribution} \ee

\indent
The feedforward connections are denoted by the matrix $\Jm \in 
\matreal{P}{N}$ and its matrix elements by 
$\{J_{ij}\} (i=1,...,P; j=1,...,N)$. Instead of considering a fixed 
matrix we prefer to compute the average MI over an ensemble of 
stochastic binary channels.
The $\{ J_{ij} \}$ have also a
Gaussian distribution, with zero mean value  
and two-point correlations $\Gammam$:

\be
\ll J_{ij} J_{i^{\prime}j^{\prime}} \gg = 
\delta _{ii^{\prime}} \Gamma_{jj^{\prime}},
\label{J}
\ee 

\noindent
where the double angular brackets indicate the average over 
the channel ensemble and $\Gammam \in \matreal{N}{N}$ .
Notice that those connections converging to different outputs are 
independently distributed.
The coupling matrix $\Jm$ can be also regarded as $P$ random  $N$-dimensional 
vectors $\vec{J_{i}} (i=1,...,P)$ given by the rows of \Jm. From 
eq.(\ref{J}), we have that each of these rows is distributed 
independently as~:

\be \rho (\vec{J_{i}} ) = \frac{e^{- \frac{1}{2} \vec{J_{i}} (\Gammam 
)^{-1} \vec{J_{i}^{t}}  }}{\sqrt{\det{(2 \pi \Gammam )}}}\; . 
\label{Jdistribution}\ee

\indent
Let us now define the local field as $\hv = \Jm\:\xiv$. 
The output state is computed by means of the probability 
distribution $\prob (\vv \, | \hv \, )$ that the output 
vector is $\vv$ for a  given  local field  $\hv$. 
Most interesting problems have a factorized output 
distribution, we will then assume that 
$\prob (\vv\, |\hv\, ) = \prod_{i=1}^{P} \prob (v_{i} | h_{i})$.
We will also require the reasonable  condition that 
$\prob (v_{i} | h_{i} ) = f_{\beta}( h_{i} v_{i} )$
($\beta$ denotes an output noise parameter), where $f_{\beta}(x)$ is an 
arbitrary function satisfying $0\leq f_{\beta}(x)<1$ and 
$f_{\beta}(x)+f_{\beta}(-x)=1$.

\indent
Both the replica and the exact calculations will be done for 
any of those $f(x)$. However we will study with some detail 
the case (to be referred to as the Hyperbolic Tangent Transfer  
function, HTT):

\be f(x)= \frac{e^{\beta x}}{e^{\beta x} + e^{- \beta x} }\;
= \frac{1}{2} (1 + \tanh{(\beta x )} )\; . 
\label{HTT} 
\ee

\indent
The deterministic channel \cite{NP1} is obtained either when $f(x)$ 
is chosen as the Heaviside function $\theta$ or in the large $\beta $
limit of the HTT function.

\indent
Let us now define the mutual information 
$I(\vv ,\xiv|\Jm )$ \cite{Shannon,Blahut,Cover} between the input 
and output vectors, given the channel parameters $\Jm$:
\footnote{In eq. \ref{elka1} and hereafter 
$\log(x) \equiv \frac{\ln (x)}{\ln 2}$. In the derivations, however, we 
make use of $\ln(x)$ because of the Taylor expansions.}

\be
I(\vv ,\xiv | \Jm) = \sum_{\vv ,\xiv } \prob(\vv ,\xiv |\Jm) 
\log \frac{\prob(\vv ,\xiv |\Jm)}{\prob(\vv |\Jm) \rho(\xiv)} \; , 
\label{elka1}
\ee

\noindent
where $\prob(\vv |\Jm)$ is the output vector distribution given $\Jm$. 
The joint probability $\prob(\vv,\xiv|\Jm )$ can be written as

\be
\label{elka2} 
\prob(\vv,\xiv|\Jm ) = \rho (\xiv) \prob(\vv | \xiv , \Jm) ,
\ee

\noindent 
where $\prob(\vv | \xiv , \Jm)$ denotes the conditional distribution 
of the output vector $\vv$ given the input $\xiv$ and the channel $\Jm$. 
Since the relation between the input and the local field $\hv$ is deterministic,  
it can be substituted by $\prob(\vv | \hv)$.

\indent
First we need to define the  output entropy for fixed couplings $\Jm $ 

\be  H(\vv | \Jm)= - \sum_{\vv} \prob (\vv | \Jm) \log{ \prob (\vv | \Jm) } , 
\label{output_entropy}
\ee

\noindent
and the entropy of the output conditioned by the input $\xiv $, 
again for fixed couplings (the equivocation term):

\be 
H(\vv |\xiv ,\Jm ) = -\sint{N}{\xiv } \rho (\xiv ) \sum_{\vv } \prob 
(\vv | \xiv ,\Jm ) \log{\prob (\vv | \xiv ,\Jm )} \; .
\ee

\noindent
Then the MI can be expressed as

\be
I(\vv ,\xiv |\Jm )= H(\vv |\Jm ) - H(\vv |\xiv ,\Jm ). 
\ee

\noindent
We are interested in the mutual information $I= \ll I(\vv ,\xiv|\Jm ) 
\gg$ averaged over the channel ensemble. Then,  calling $I_{1} = \ll H(\vv 
|\Jm ) \gg$ and $I_{2} = \ll H(\vv | \xiv ,\Jm ) \gg$, we have $I= I_{1} - 
I_{2}$, or in terms of MI per input unit, $\ii = \ii_1 -\ii_2$, where
$\ii_1=\frac{I_1}{N}$ and $\ii_2=\frac{I_2}{N}$.

\indent
Each term will be studied separately. In the next section we compute 
the equivocation term $I_2$. This can be obtained exactly by means of simple 
arguments. The output entropy term requires more care. It will be 
evaluated in section \ref{I1RSA} using the replica technique and 
in section \ref{analytical} using exact analytical methods.

\section{The equivocation term $I_{2}$}
\label{I2_presentation}

\indent

Since $\prob (\vv |\xiv ,\Jm )$ factorizes, 
it is convenient to define the single output entropy, $\bitent{h_{i}}$, 
which is a {\em function} of $h_{i}$~:

\be \bitent{h_{i}} = - \sum_{v_{i}=\pm 1 } \prob (v_{i} | h_{i})
\ln{\prob (v_{i} | h_{i})}\; , 
\ee

\noindent
where one should keep in mind that $\hv = \Jm \xiv$. Then~:

\be H(\vv | \xiv ,\Jm )= \sum_{i=1}^{P} \sint{N}{\xiv } \rho (\xiv ) 
\bitent{h_{i}}   \ee 

\noindent
and~

\be 
I_{2}= \sum_{i=1}^{P} \sint{}{\Jm} \rho (\Jm ) \sint{N}{\xiv } 
\rho (\xiv ) \bitent{h_{i}} . 
\label{I2_naif}
\ee 

\indent
$I_2$ can be easily evaluated in several simple examples. 
In the deterministic case it is zero. In the large noise limit ($\beta 
\rightarrow 0$) it reaches its upper bound $I_{2}=P \ln{2}$. 
For the HTT function, eq.~(\ref{HTT}), we have $\bitent{h} = 
\ln{(e^{\beta h}+e^{- \beta h})} - \beta h \:\tanh{\beta h}$,  and this 
single output entropy can be substituted in eq.~(\ref{I2_naif}) to obtain 
$I_2$. In Appendix \ref{I2} the details of such calculation are 
presented; here we recall the final formula eq.~(\ref{generalI2}), valid 
for a matrix $\Mm=\Gammam \Gm$ having all its eigenvalues of the same order
and any function $\bitent{h}$. The equivocation term {\it per} input unit 
is then:

\be 
\ii_{2}= \alfa \int_{- \infty}^{\infty} dz\; \frac{e^{-z^{2}}}{\sqrt{\pi}} 
\;\bitent{\sqrt{2 \bM }\, z} \; .
\label{I2piu}
\ee

\noindent
with $\bM=$ Tr(\Mm ) (the trace of $\Mm$). For the sigmoidal HTT 
function, we obtain:

\be 
\ii_{2}= \alfa \sum_{m=1}^{\infty} (-1)^{m+1} A_{m} ,
\ee

\noindent
where

\be A_{m} = \frac{2 \beta_0}{\sqrt{ \pi}} + (\frac{1}{m} - 2 m \beta_0^{2})
(1 - \erf{ m \beta_0 }) e^{ m^{2} \beta_0 ^{2} }
\ee

\noindent
and $\beta_0 = \sqrt{2 \bM }\, \beta$, which we shall call the {\it reduced 
noise parameter\/}. Alternatively, we will also use the {\it reduced 
temperature} $T_0 = {\beta_0}^{-1}$. The symbol $\erf{x}$ stands for the 
error function. 

\indent
One can easily obtain several limits. For small $T_0 $~:

\be \ii_{2} \approx   \alfa \frac{\pi ^{3/2}}{6} T_0 
\label{smallTI2} \; ,
\ee 

\noindent
where one observes that  $i_{2}$ grows linearly with $T_0$ .
For small $\beta_0 $,

\be \ii_{2} \approx \alfa\:\ln{2} - \frac{\alfa}{4}\:\beta_0 ^{2}\; ,
\label{largeTI2}
\ee 

\noindent
in this case $\ii_{2}$ departs from $\alfa \ln{2} $ quadratically with 
$\beta_0 $ .

\indent
For other transfer functions one obtains the same qualitative result, 
although with different coefficients. This is simply because these 
coefficients depend on the derivatives of the transfer function in the 
neighborhood of $\beta=0$ and $\beta=\infty$, respectively.

\section{Calculation of $I_{1}$} 
\label{I1}

\indent

We now compute the output entropy term defined in eq.(\ref{output_entropy}).
First we notice that the (discrete) probability density of the 
outputs $\vv$ is given by:

\be \prob _{\vv } \equiv \prob (\vv|\Jm) =  
 \sint{N}{\xiv } \rho (\xiv ) \prob (\vv |\xiv, \Jm )\; . 
\ee 

\noindent
Since $\prob_{\vv}$  is a probability density, $\sum_{\vv } \prob_{\vv }
=1$ and so

\be I_{1} = -\lim_{n \rightarrow 0}\: \frac{\sum_{\vv } \ll \prob 
^{n+1}_{\vv } \gg\,\, -1}{n} \; . \ee

\noindent 
$\ll\prob _{\vv }^{n+1} \gg $ can be written in terms of replicated
variables $\{\xiv^{\ai} \} $,  $\ai = 0,1,...,n$ in the following way~:

\be 
\ll\prob _{\vv }^{n+1} \gg = \sint{}{\Jm } \rho(\Jm) 
\int \prod_{\ai =0}^{n}
\left[d^{N}\!\xiv^{\ai} \rho _{\xiv }(\xiv^a ) 
\prod_{i=1}^{P} f( v_{i} h_i^{\ai} ) \right]\; .
\label{l1}
\ee 

\noindent
Here the local fields are:

\be h_i^{\ai}  = \sum_{j=1}^{N} J_{ij}\: \xi_j^{\ai} \; . \ee 

\indent
We will only compute the integer order moments of $\prob _{\vv }$.  
The  continuous order moments 
will be obtained by naive extrapolation of them. 
Actually they can be obtained in a completely 
rigorous way (although in a rather complicated fashion) because the integer 
moments contain enough information to reconstruct 
the probability distribution of the variable $\prob _{\vv }$.

\indent
It is obvious from  the distribution of $\Jm $ that each element $J_{ij}$ is 
independent of the others with different output index $i$. Using this 
and the fact that each $\vec{J_{i}}$ has an even distribution  
(Gaussian) we have that  $\ll \prob _{\vv }^{n+1} \gg$ is 
independent of $\vv $. Thus we can write~:

\be I_{1} = - \frac{2^{P} \ll \prob^{n+1} \gg\,\, -1}{n} ,\,\, n 
\rightarrow 0 \; , \label{limitI1} \ee 

\noindent
where $\prob $ stands for the conditional distribution $\prob _{\vv } $ 
with a specific choice of $\vv $ , e.g. $v_{i}=+1, i=1,...,P $.

\indent
In eq.~(\ref{l1}) we can apply, for almost every $\Jm $, the
Bessel-Plancherel identity in each of the integrals over the
$\xiv^{\ai}$'s:

\be 
\label{l2}
\sint{N}{\xiv^{\ai}} \frac{e^{- \frac{1}{2} \xiv^{\ai t}
( \Gm )^{-1} \xiv^{\ai } } } {\sqrt{\det{(2 \pi \Gm )}}} 
\prod_{i=1}^{P}
f(h_{i}^{\ai}) = \sint{P}{\uv^{\ai}} e^{- 2 \pi ^{2} \uv^{\ai t} 
\Deltam \uv^{\ai} } \prod_{i=1}^{P} \hat{f} ( 
u_i^{\ai}) \; .\ee 

\noindent
The function $\hat{f}$ is the Fourier transform of $f$ and must be 
understood in the distributional sense, and $\Deltam = 
\Jm\,\Gm\,\Jm ^{t}$. This expression holds for any
value of $P$ and $N$ (see Appendix \ref{PgreaterN} for details). Then,
the characteristic function  $\hr{\{ \uv^{\ai} \} _{\ai
=0,...,n}}$ associated to the joint probability distribution of the
replicated local fields, $\rho (\{ \hv^{\ai} \}_{\ai=0,...,n})$ is:

\be \hr{\{ \uv^{\ai} \} _{all\:\ai 's} }= e^{- \frac{1}{2} 
\sum_{j=1}^{N} \ln{\det{(\id{P} + 4 \pi ^{2} m_{j} \Um)}} } ,
\label{hatrho_1}
\ee

\noindent 
where the matrix $\Um $  is the sum of all the projectors associated to each 
replica vector,

\be U_{ii'} = \sum_{\ai=0}^{n} u_i^{\ai} u_{i'}^{\ai}. 
\ee

\indent
 From eqs.(\ref{l1}) and (\ref{l2}) we obtain the following result:

\be \ll \prob ^{n+1} \gg = \int \prod_{\ai = 0}^{n} d^{P} \!\uv^{\ai} 
\hr{ \{ \uv^{\ai} \} } \prod_{\ai = 0}^{n} 
\prod_{i=1}^{P} \hat{f} (u_{i}^{\ai}) 
\label{startmethod2}\; . 
\label{momentprob}
\ee

\noindent
This formula is exact for the moments of integer order of \prob .
This will be the starting point of subsection \ref{analytical}, where we
will calculate the moments of $\prob$ exactly. Before that we present the 
RSA solution in the next subsection.

\subsection{The RSA approach}
\label{I1RSA}

\indent

After a rather lengthy algebra we obtain the entropy
term $I_{1}$ . Some details of the calculation are 
described in Appendix~\ref{RSAapp}; here we only give the final result. 
$I_{1}$ is a function of two order parameters, here called $x$ 
and $s$. Its value per input unit is given by:

\be \ii_1^{RSA} = \alfa R(x) + \frac{1}{2} \tau [\:\ln{(\id{N} +  s \Ggm 
)}\: ] -\frac{s}{2 (x+1) } \label{RSAI1}\; , \ee 

\noindent
where  $\Ggm $ is the normalized matrix, $\Ggm = \frac{ N \Mm }{\bM} $.
We have made use of the symbol $\tau $ for a normalized trace 
operator, $ \tau(\:\cdot\: ) = \frac{1}{N} Tr (\:\cdot\; )$. The order 
parameters satisfy the self-consistent SP equations:

\be \left\{ \begin{array}{ccc}
x & = & s \tau ( \frac{\Ggm ^{2}}{\id{N} + s \Ggm}) /
\tau ( \frac{\Ggm }{\id{N} + s \Ggm } ) \\
s & = & -2 \alfa (x+1) ^{2} \frac{dR}{dx}
\end{array} \right. \label{SPeqs} \; ,\ee 

\noindent
The function $R(x)$ is the average entropy of an effective transfer 
function $g_x$ defined as:

\be g_x(y) = \frac{1}{\sqrt{\pi }}  \int_{-\infty}^{\infty} 
dw \; e^{ -(w + y)^{2} } f(\sqrt{\frac{ 2 \bM}{1+x}} w)\; . \ee 

\noindent
More precisely,

\be R(x) = \frac{1}{\sqrt{\pi }} \int_{-\infty}^{\infty}  dy\; e^{-y^{2}}
S( g_x(\sqrt{x}\, y) )\; , \label{averageentropy} \ee 

\noindent
where $S(z)$ is the entropy of a binary probability, i.e. ,

\be S(z) = - z \ln{z} -(1-z) \ln{(1-z)}\; . 
\label{bitentropy}
\ee 

\indent
It is worth noting that for the deterministic case $g_x(y) =\frac{1}{2} 
(1+\erf{y}) $, and for the fully random case $g_x(y) = \frac{1}{2}$.
In the deterministic case, there is a simple relation between our
parameters and those ($q$ and $\hat{q}$)used in \cite{NP1}: $\hat{q} = s $ 
and $q = \frac{x}{1+x} $. We prefer to use $x$ instead of 
$q$ because it usually yields simpler expressions.

\subsection{The Exact Solution}
\label{analytical}

\indent

In this subsection we present an exact evaluation of $I_1$ valid for
$\alfa\leq\alfa_c$, where $\alfa_c$ is of order one.
It is necessary to assume that all the eigenvalues of $\Mm=\Gammam \Gm$ 
are of the same order. The details of the calculation are presented 
in Appendix \ref{Analyticapp}; we only give here the final result for the 
moments:

\ba \ll\prob^{n+1}\gg = 2^{-P (n+1) } \; e^{-\frac{1}{2} Tr[\:\ln{(\id{N} 
-\frac{2}{\pi}\: k^{2}\: n\: \alfa\:\Ggm )}\: ] }
\nonumber \\
e^{-\frac{n}{2} Tr[\:\ln{(\id{N} +\frac{2}{\pi}\: k^{2}\: \alfa\:\Ggm 
)}\: ] } 
\label{moments}\; , \ea

\noindent
where $k$ is defined as:

\be  k = \int_{-\infty}^{\infty} dy\;\; y\, e^{-\frac{y^{2}}{2}} 
f(\sqrt{\bM}\; y) \; .\ee 

\noindent
Let us now extrapolate eq.~(\ref{moments}) to non-integer $n$. This gives
our analytical estimate of $\ii_{1}$~:

\ba \ii_{1}^{an} = -\,\frac{1}{N}\, \lim_{n\rightarrow 0}\,
\frac{2^{-P (n+1)}\ll\prob^{n+1}\gg\; -1}{n}
\nonumber \\ 
=\; \alfa\ln{2} -\frac{k^{2}}{\pi}\:\alfa
+\frac{1}{2} \tau[\:\ln{(\id{N}+\frac{2 k^{2}}{\pi}
\:\alfa\:\Ggm)}\: ] 
\;  .\label{analyticI1}\ea

\noindent
The numerical constant $k$ is real-valued, and it is expected to be of
order one. In fact, the 
maximum of $k$ is reached in the deterministic case ( $f(x)=\theta(x) $ ). 
This gives $k=1$, independent of $\bM $. Notice that the 
minimum of $k$ is reached by $f(x)=\theta (-x)$, giving $k=-1$. The 
minimum of $k^{2}$ is realized by $f(x)=\frac{1}{2}$, what gives $k=0$.

\indent
The computation done in this section (eq.~(\ref{moments})) is
equivalent to a Taylor expansion of the original equation for the
moments, eq.~(\ref{startmethod2}). 
This can be checked by explicit
evaluation of the derivatives of the two expressions.\footnote{
To compute the derivatives of eq.~(\ref{startmethod2}) 
with respect to $\alfa$ one has first to make explicit its dependence
on the parameter N by expressing $\Mm$ in terms of $\Ggm $ 
($\Mm = \frac{\bM}{N} \Ggm$).  
Then, after setting each derivative at $\alfa = 0$, 
the resulting integrals are easy to compute.
}

\section
{Analysis of the RSA solution}
\label{expansions}

\indent

Using eq.~(\ref{RSAI1}), together with the SP equations~(\ref{SPeqs}), we  
obtain expansions of $\ii_1^{RSA}$ at small and large $\alfa $ and 
$\beta_0$ ($\beta_0= \sqrt{2\bM}\beta $). 
We will make explicit calculations for the 
deterministic and the HTT functions (the completely random case 
always gives  $\ii_{1} = \alfa \ln{2}$).

\subsection{Small $\alpha $ limit}

\indent

Let us first investigate the deterministic case ($\beta_0 \rightarrow 
\infty$) in this regime. From eq.~(\ref{SPeqs}), we can see that  $s 
\approx \frac{2}{\pi }\alfa$ and $x 
\approx s\: \tau(\Ggm ^{2}) \,\,\approx \frac{2 \tau(\Ggm ^{2})}{\pi 
}\,\alfa$. This gives the first two orders of the expansion of $\ii_1$ in 
powers of $\alfa$:

\be \ii_1^{RSA} \sapprox{\alfa\ll 1} \alfa \ln{2} - \alfa ^{2} 
\frac{\tau(\Ggm ^{2})}{\pi ^{2} } \label{sa} \; ,\ee 

\noindent
where we see that, as expected, the second order is negative. The next 
order in $T_0$ gives, after solving the SP equations up to order $T_0$,

\be \ii_1^{RSA} \lapprox{\alfa\ll 1,T_0 \ll 1} \alfa\ln{2} - \alfa ^{2} 
\frac{\tau(\Ggm^{2})}{\pi ^{2}} +  \alfa^{2} \frac{\tau(\Ggm ^{2})}{3} T_0^{2} 
\label{sasT} \; . \ee

\noindent
This is a positive contribution. However, this does not mean that the MI 
increases with $T_0$; in fact, the term $\ii_2$ gives a larger contribution
of order $\alfa T_0$, as can be seen in eq.~(\ref{smallTI2}). More precisely,

\be \ii^{RSA}\: =\: \ii_1^{RSA}-\ii_2\: \approx \alfa \ln{2} -  \frac{\pi 
^{3/2}}{6} \alfa T_0  - \alfa ^{2} 
\frac{\tau(\Ggm ^{2})}{\pi ^{2}}
+ \frac{\tau(\Ggm ^{2})}{3} \alfa ^{2} 
T_0^{2}\; . \ee 

\indent
We now calculate the first order correction in $\beta_0 \ll 1 $ to 
$\ii_1^{RSA}$. The leading order is the fully stochastic case, and
$\ii_1^{RSA} = \ii_{2} = \alfa \ln{2}$ ($x = s =0$). Up to the next order, 
$\ii_1^{RSA}$ is:

\be \ii_1^{RSA} \lapprox{\alfa\ll 1, T_0\gg 1} \alfa\ln{2} - \alfa ^{2} 
\frac{\tau(\Ggm ^{2}) }{16}  \beta_0 ^{4} \label{salT} .\ee 

\noindent
 From eqs.~(\ref{largeTI2}) and (\ref{salT}), we obtain~:

\be \ii^{RSA} \approx \frac{1}{4} \alfa \beta_0 ^{2} - \frac{\tau(\Ggm 
^{2})}{16} \alfa ^{2} \beta_0 ^{4} . \ee

\subsection{Large $\alpha $ limit}

\indent

In this regime, and in the low temperature limit, we obtain from 
eq.~(\ref{SPeqs}) that $x\approx s$ and $s\approx A_0 \alfa\sqrt{x}$, 
where $A_0$ is the constant given in \cite{NP1}, \cite{KNP}~:

\noindent
$A_0=\frac{1}{\sqrt{\pi }} \int_{-\infty}^{\infty} 
dz\; e^{-z^{2}} S(\frac{1+\erf{z}}{2})$, $A_0\approx 0.72$. From here 
we obtain $s\approx A_0^{2} \alfa ^{2} $, $x\approx A_0^{2} \alfa ^{2}$. 
Substituting these parameters in eq.~(\ref{RSAI1}) one obtains the known 
result  \cite{NP1}, \cite{KNP}:

\be \ii_{1}^{RSA} \sapprox{\alfa\gg 1} \ln{\alfa } +\frac{1}{2} + \ln{A_0}
+ \frac{1}{2} \tau [\:\ln{\Ggm }\: ] \label{la}\; . \ee

\noindent
Adding weak output noise, and assuming $\alfa\: T_0\rightarrow 0$ we have:

\be \ii_{1}^{RSA} \lapprox{\alfa\gg 1,T_0\ll 1} \ln{\alfa }+ \frac{1}{2} +  
\ln{A_{0}} + \frac{1}{2} \tau[\:\ln{\Ggm}\: ] +  \frac{\pi ^{2} 
A_{0}^{2}}{12} \alfa^{2} T_0^{2} \; .\ee 

\noindent
 From here and eq.~(\ref{smallTI2}):

\be \ii^{RSA} \approx \ln{\alfa } +\ln{A_0} +\frac{1}{2}+ 
\frac{1}{2}\tau[\:\ln{\Ggm}\: ] 
-\frac{\pi ^{3/2}}{6} \alfa T_0 
+\frac{\pi ^{2} A_0^{2}}{12} \alfa ^{2} T_0^{2}\; . \ee

\noindent
In the opposite limit, $\beta_0 \ll 1$ (large temperatures), and also 
assuming $\alfa \beta_0^2$ small, it is straightforward to see that~:

\be \ii_{1}^{RSA}\lapprox{\alfa\gg 1, T_0\gg 1} \alfa \ln{2} - \frac{\beta_0^{2}}
{4} \alfa +\frac{1}{2} \tau[\:\ln{(\id{N} +\frac{1}{2} \beta_0 
^{2}\alfa\Ggm )}] \label{lalT} \; ,\ee 

\noindent
which together with eq.~(\ref{largeTI2}) gives:

\be \ii^{RSA} \approx \frac{1}{2} \tau[\:\ln{(\id{N} +\frac{1}{2}\beta_0 
^{2}\alfa\Ggm )}\: ]\; . \ee 

\noindent
which shows that the MI decays as $\beta_0 ^{2}$ when $\beta_0\rightarrow 0$.

\subsection{Numerical analysis}

\begin{figure}[htb]
\begin{center}
\hbox{
        \makebox[.5cm]{$\ln \protect\ii^{RSA}$}
        \makebox[12cm]{}
}
\vspace*{0.1cm}
\hbox{
        \makebox[.5cm]{}
        \makebox[12cm]{
		\leavevmode
		\epsfxsize=12cm
		\epsfbox[50 50 410 302]{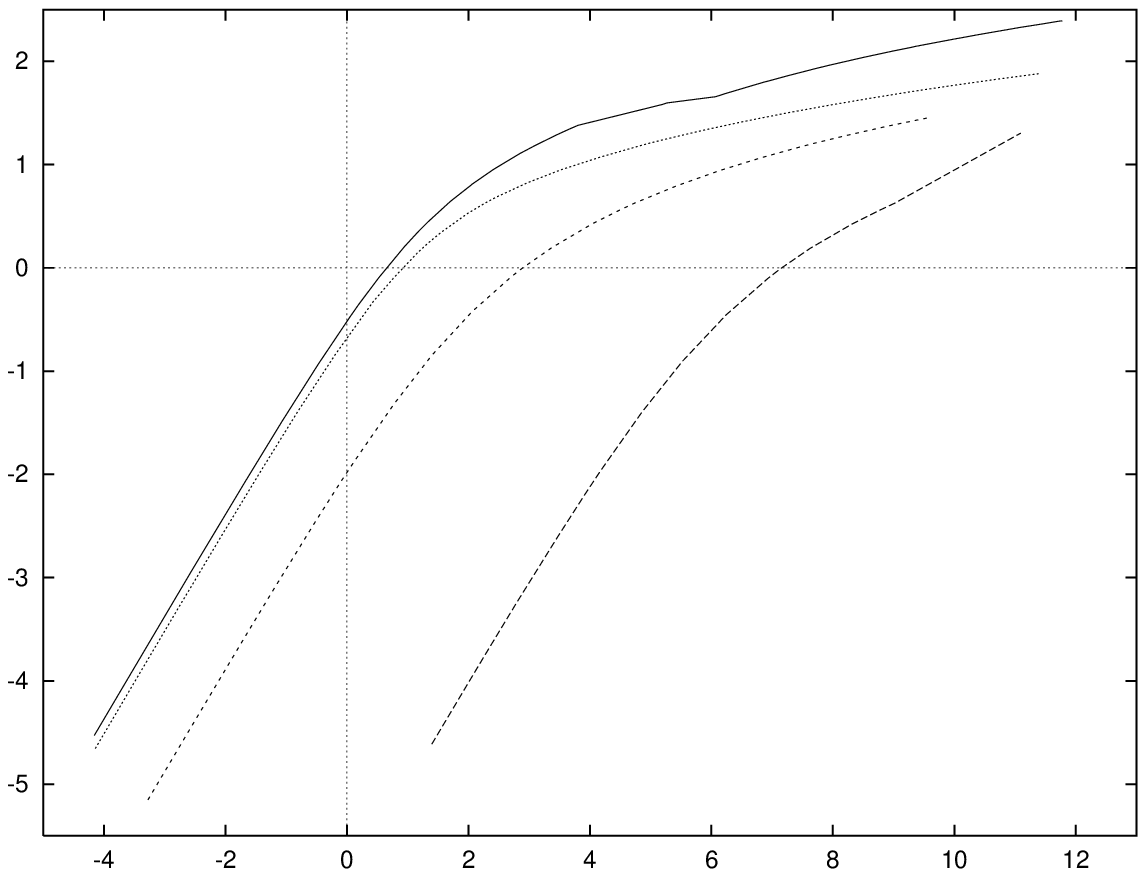}
	}
}
$\ln\protect\alfa$
\end{center}
\caption{ $\protect\ln \protect\ii^{RSA}$ vs. $\protect\ln\protect\alfa$
computed with the RSA for the deterministic transfer function and the
HTT function, for several values of $\protect\beta_0$ : 
\protect\newline
1. Solid line: $\protect\beta_0 =\protect\infty$
(Deterministic)\protect\newline
2. Dotted line: $\protect\beta_0 = 10$ (near to
deterministic)\protect\newline
3. Light dashed line: $\protect\beta_0 = 1$\protect\newline
4. Dashed line: $\protect\beta_0 =0.1$ (not far from full 
stochasticity) 
}
\label{variasTs}
\end{figure}

\noindent
The plot of~~$\ln\ii^{RSA}$  (which is obtained combining 
eqs.~(\ref{I2piu}) and (\ref{RSAI1})) versus $\ln{\alfa }$, using 
the HTT for several values of the reduced noise parameter  $\beta_0 
=\sqrt{2\bM } \; \beta $, is shown in Fig.~\ref{variasTs}. The correlation  
matrix was taken proportional to the identity: $\Mm = \frac{\bM}{N} 
\id{N} $ . As expected, for each $\alfa$, the MI
decreases as the temperature increases. It is also interesting that an 
increase in the temperature moves the saturation point 
(the change from the close-to-linear regime to the asymptotic one, in 
which the MI increases slower with $\alfa$) to greater values of $\alfa$.

\section{Comparison between the exact and the RSA solutions}
\label{comparison}

\indent

The analytical result presented in eq.~(\ref{analyticI1}) seems rather 
astonishing as it provides a very simple expression for $\ii_{1}$, 
compared with the cumbersome formulae of the RSA solution. Then the 
following two questions arise: first, whether the two
solutions do or do not coincide at least in the range of 
validity of the exact one. Secondly, if the exact MI can be analytically 
extended to greater values of $\alfa$. We will see that
the answer to both questions is no, at least for the deterministic 
transfer function.

\begin{figure}[htb]
\begin{center}
\hbox{
        \makebox[.5cm]{$\ln\protect\ii$}
        \makebox[12cm]{}
}
\vspace*{0.1cm}
\hbox{
        \makebox[.5cm]{}
        \makebox[12cm]{
		\leavevmode
		\epsfxsize=12cm
		\epsfbox[50 50 410 302]{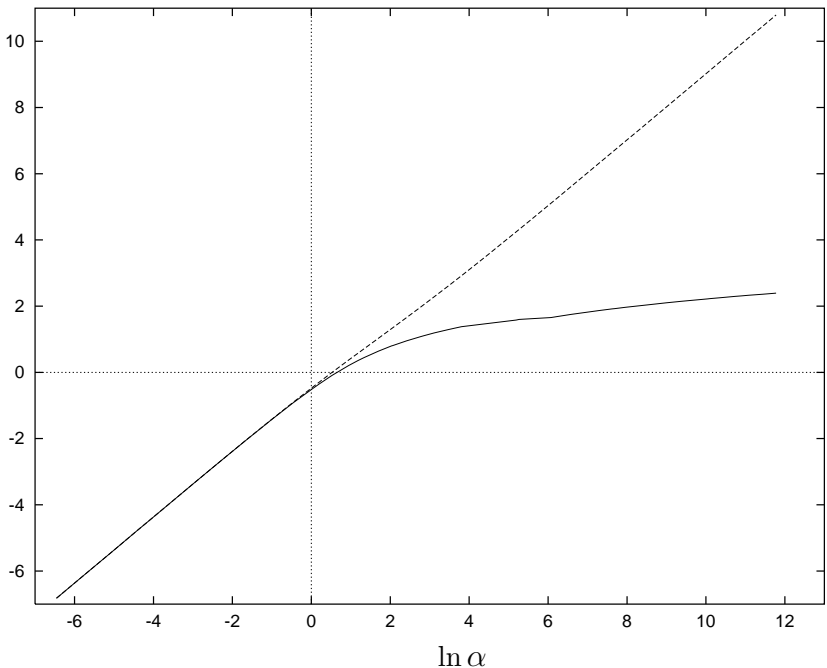}
	}
}
$\ln\protect\alfa$
\end{center}
\caption{ $\protect\ln\protect\ii$ vs. $\protect\ln\protect\alfa$
for the RSA solution (solid line) and 
for the analytical solution (dashed line), for the deterministic transfer 
function.
} 
\label{divergencia}
\end{figure}

\indent
With respect to the first question, an expansion in powers of $\alfa$ can 
be easily evaluated for the deterministic case. It turns out that the 
corresponding Taylor coefficients coincide up to the second order, but the 
third is different. For instance, if the matrix $\Mm$ is 
proportional to the identity we observe that $\ii^{\: an} - 
\ii^{RSA}=\ii_1^{\: an} - \ii_1^{RSA}\approx 
\frac{4}{\pi^{4}} \alfa^{3}$ at the lowest order in $\alfa$. It should be 
noted that $\ii^{an}$ is always greater than $\ii^{RSA}$
(see figure \ref{divergencia}). Both graphs are very close up to an 
undetermined value  of $\alfa $ near $\alfa =1$ ( $\ln\alfa\approx 0$ ), 
from which they split away fast. Detailed numerical 
studies for small $\alfa $ ( $\alfa\in [ .0001 \, ,\, .005 ]$ ) 
confirmed a cubic divergence between the two MI's with coefficient  
$\approx\frac{4}{\pi^{4}}$ and a deviation form this value 
less than $ 0 .25\,\% $.

\indent
As to the second question, the large $\alfa$ expansion of the RSA 
solution is (eq.~(\ref{la})):

\[
\ii_{1}^{RSA} \sapprox{\alfa\gg 1} \ln{\alfa } + \frac{1}{2}
+ \ln{A_0} +\frac{1}{2} \tau[\:\ln{\Ggm }\: ] \; . 
\]

\noindent
It is consistent with what is known about the continuous outputs, 
which should be reproduced when $\alfa$ goes to infinity. 
On the other hand,  the analytical 
solution gives, in this limit:

\be \ii_{1}^{\: an} \sapprox{\alfa\gg 1} \alfa\left(\ln{2} 
-\frac{1}{\pi}\right) +\frac{1}{2} \ln{\alfa } + 
\frac{1}{2}\ln{\frac{2}{\pi }} + \frac{1}{2} \tau[\:\ln{\Ggm }\: ] ,\ee

\noindent
which is a qualitatively very different behaviour.  Since the
analytical solution is exact for small $\alfa$, with a convergence
radius $O(1)$, the previous expansion suggests that the channel
exhibits a phase transition.

\section{Discussion of the phase transition}
\label{phase}

In this section we give a series of arguments to support 
the existence of a phase transition at $\alfa_c=O(1)$.

\begin{enumerate}

\item
The first argument is provided by the  
behavior of the moments. The analytical computation of the integer 
moments, eq.~(\ref{moments}), is exact in the thermodynamic limit.
Yet, those moments cannot be correct for every value of 
$\alfa$. This is because they diverge at the values $\alfa_c^n =  
 \pi / ( 2 k^{2} n) $. On the other hand 
since $\prob $ is a positive variable bounded by one (and then its 
moments should be less than one) one can conclude that eq.~(\ref{startmethod2}) 
presents critical points before those values. 
A natural guess would be that these singularities appear 
at values of $\alfa$ that follow the same behavior $1 / (k^{2} n)$.
\footnote{
{\bf Remark:\/}  Since the moments
factorize as the product of contributions related to each eigenvalue 
of \Mm , we expect that there is a critical value of 
$\alfa $ for each of them. The functional behaviour at these transitions is 
the same, differing only in the critical value of $\alfa $ 
where they occur. This is not a serious complication, although one sould 
keep in mind that the distribution of eigenvalues of $\Mm$ is relevant.
}

\item
The critical point of $\ll\prob\ln{\prob}\gg$  is related 
to the critical points of the moments (\ref{startmethod2}).  
We then expect that it has a phase transition at some 
$\alfa_c \sim  1/k^{2} $. 
As an example we consider the completely random channel 
($k = 0$), where according to the previous argument 
the transition in pushed to infinity. 
In fact, the expansions of $\ii_{1}$ computed with the analytical and the 
RSA solutions coincide in the large T limit, in both regimes of $\alfa $ 
(eqs. (\ref{salT}) and (\ref{lalT})).

\item
One could infer the existence of the critical point, observing the 
behaviour of the probability density $\rho(\vec{h})$.
If one considers this distribution for only one replica, a 
dramatical change in the shape of the function takes place  when $\alfa 
$ goes from 1 to 2. Considering its Fourier transform, eq.~(\ref{l4}), it 
can be seen that $\hr{\uv}$ behaves at $\infty$ like $u^{-N}$ 
($u$ denotes the modulus of $\vec{u}$) , while the 
volume element behaves as $u^P$. This means that this function is 
integrable (i.e., $\hr{\uv}$ is a $L^{1}$ function\footnote{
$L^q=\{f: \left[\int|f|^q \right]^{\frac{1}{q}} < + \infty\}$
}
)
up to ${\bf \alpha =1}$. It is also a square integrable function
(i.e., it belongs to $L^{2}$) in this range. From $\alfa =1$ to 
$\alfa =2$ it is no longer a $L^{1}$ function, but it still belongs to
$L^{2}$; and beyond $\alfa = 2$ it is no longer in $L^{2}$. 
What does this mean in terms of $\rho(\hv)$ ?

\begin{itemize}
\item
{\bf Below} ${\bf \alpha =1}$, $\hr{\uv}\;\in\;L^{1}$. Consequently its 
Fourier transform $\rho(\hv)$ is bounded (that is, belongs to 
$L^{\infty}$). Besides, since $\rho(\hv)$ is a probability density it 
is also in $L^1$. The same argument holds for its derivatives in the 
thermodynamic limit. This is because derivation in $\hv$-space is 
equivalent to multiplication 
by powers of $\uv$ in $\uv$-space. Since the order of the derivatives is 
finite, the leading behaviour in the thermodynamic limit is not changed.

\indent
It follows that $\rho(\hv)$ and all its derivatives belong to
$L^{1}\cap L^{\infty}$. This means that $\rho(\hv)$ belongs to the 
Schwartz's class (see for instance \cite{Rudin}). Then, it is a very 
regular, fast decreasing function.

\item
{\bf Beyond} ${\bf \alpha =2}$, $\hat{\rho }$ is no longer in 
$L^{2}$, so $\rho $ cannot belong to $L^{2}$ either (since the Fourier 
transform is an isomorphism on $L^{2}$). Thus, $\rho $ 
cannot belong to $L^{\infty }$ (as $\rho $ belongs to $L^{1}$, then 
it would belong to $L^{1}\cap L^{\infty }\; \subset L^{2}$, which is 
a contradiction~). Then the graph of $\rho $ is broken by one or more 
divergences to $\infty $.

\item
{\bf Between ${\bf \alpha =1}$ and ${\bf \alfa =2}$}, the transition between 
the other two regimes has to occur.

\end{itemize}

\indent
For more than one replica, heuristic arguments permit to 
say that the main contributions to the characteristic function behaves like
$u^{-N}$, independent of the number of replicas. 
The volume element behaves like $u^{P(n+1)}$. By the same arguments used 
in the case of a single replica, now 
$\rho(\{\vec{h}^{\ai}\}_{\ai=0}^n)$
exhibits a transition which takes place between $\alfa = 1/(n+1)$ and 
$\alfa = 2/(n+1)$. This is in agreement with the main conclusion 
obtained in the first comment.

\indent
Thus, we have proved that the joint probability distribution of the 
replicated fields, $\rho(\{\vec{h}^{\ai}\}_{\ai=0}^n)$, undergoes a 
phase transition at a some finite $\alfa$.
Recalling that

$\ll \prob^{n+1} \gg$ is calculated averaging
$\prod_{\ai=0}^{n} \prod_{i=1}^{P} f(h_i^{\ai})$ with 
this function, it is thus reasonable to 
think that the integer moments of $\prob$ and the MI could exhibit a phase 
transition caused by the transition in the own distribution.

\item
Another argument in favor of the existence of a transition is given by
the behavior of the information capacity.  It has been proved
\cite{NP1} that this quantity has a third order transition for the
deterministic channel.  The high order of this transition 
makes the function rather smooth and the critical point hard to detect.
The information capacity is only an upper bound of the MI, 
but it is plausible that the latter has a similar behavior.

\end{enumerate}

\indent
These comments lead us to conclude that the MI undergoes a 
phase transition. What is then the meaning of the RSA solution?
The expansion in powers of $\alfa$ of the RSA solution differs from that 
of the exact one at the third order, which is precisely the order of the 
transition for the information capacity. Besides, a detailed study of 
the RSA solution shows that the dependence
of $\ii_{1}^{RSA}$ on $\alfa $ is infinitely smooth: this solution 
exhibitis no change in its behavior. 

\indent
The conclusion is that the completely symmetric ansatz does not provide a 
wide enough family of solutions and the maximal MI is not attained by 
this ansatz. This explains why the exact solution is always above the RSA 
one. So, RSA {\it seems to be\/} a smooth regularization of the true MI. 
This would explain why it splits away from the true MI in a cubic 
way, suppossing that the latter possesses a third order transition. On the 
other hand, the behaviour at large $\alfa$ of the RSA 
solution is consistent with that of the information capacity and of the 
MI in a network with continuous output. Then, it is plausible that the
RSA provides a smoothening for MI which asymptotically 
has the correct behavior, but which masks completely the critical point.

\section{Further steps: Beyond the RSA}
\label{SB}

\indent

We have explored the possibility of breaking the replica symmetry by
modifying the ansatz for $\Utm $ and $\Vtm $ (See Appendix \ref{RSAapp}).
Our first attempt consisted in the usual RSB ansatz. After rather lenghty
calculations this led us to exactly the same solution given by the RSA. 

\indent
We also tried what can be called the Segregated Ansatz (SA), in 
which the first of the replicas is split from the other $n$. Then we assume:

\be \left\{ \begin{array}{cccc}
\tilde{U}_{00}^{0} & = & U_{0}, &  \\
\tilde{U}_{0\bi }^{0} = \tilde{U}_{\bi 0}^{0} & = & U_{1}, & \bi = 
1,...,n \\
\tilde{U}_{\ai\ai }^{0} & = & U_{2}, & \ai = 1,...,n \\
\tilde{U}_{\ai\bi }^{0} & = & U_{3}, & \forall\ai\neq\bi \;\in 
\{1,...,n\} 
\end{array} 
\right.
\ee
 
\noindent
and analogously for $\Vtm_{0}$ .\footnote{
This ansatz is justified because it splits a typical 
$ n \times n $ box from the matrices, which are $(n+1)\times
(n+1)$. This splitting allows the segregated replica to 
behave independently from the others.
}

\indent
Under this ansatz the RS solution verifies again the SP equations.
But in addition, an new infinite set of functions of $\alfa$ appeared that 
also verify the SP equations. The MI will then be given, at each $\alfa$, 
by the function providing the maximal MI. We observed that at large 
$\alfa$ this infinity of solutions contributes below the RSA. However, 
for arbitrary values of $\alfa$ the problem is too complicated to deal with.


\section{Conclusions}

\label{conclusions}

\indent
In this paper we investigated the information processing by a noisy
perceptron channel. Our network has $N$ real-valued input and $P$ binary 
output neurons, which state is determined by the joint probability 
distribution of the input and output states $P(\vv, \xiv)$.
We performed the calculation for a general continuous and
bounded transfer function, depending on a noise parameter.
Our study generalizes previous results obtained for deterministic 
channels \cite{NP1} using the replica technique. We also 
give the explicit expressions for the mutual information at different 
asymptotic regimes of the load parameter $\alfa = P/N$ and the noise 
$\beta$.

\indent
The mutual information per input unit can be decomposed in two pieces:
$i = i_1 - i_2$.  The second term, which extracts the wrong bits of
information (the equivocation), can be calculated exactly because of
the factorization of the probability. The entropic part $i_1$ is more 
difficult to compute. Here we computed it by means of the replica 
technique and analytical methods.

\indent
Our main result is that for values of $\alfa$ up to some value $O(1)$
there exists an exact solution for $i$, which we found explicitly
(eq.~(\ref{analyticI1})).  This solution is {\em different} from the
replica symmetry ansatz solution (eqs.~(\ref{RSAI1}) to 
(\ref{bitentropy})).  A numerical
computation of both solutions gives the remarkable result that they
are {\em extremely close} to each other up to $\alfa \sim 1$
(Fig.~\ref{divergencia}). A small $\alfa$ expansion shows that the two 
solutions are equal up to the second order. Although the corresponding 
Taylor expansions differ above the third order, the
numerical agreement up to $\alfa \sim 1$ is excellent (a relative
difference of less than $0.9 \%$ up to $\alfa=0.1$). This is due to 
intriguing cancellations between higher orders.

\indent
Our conclusion is that there exists a critical value $\alfa_c$ of
order one, above which a drastic change of the mutual information
occurs. This signals the appearance of a phase transition.  Above
$\alfa_c$ the analytical solution is not valid, one of the reasons is
that it does not have the correct large $\alfa$ behaviour (it violates
a bound given by the information capacity).  On the other hand, even
if the replica symmetric solution is wrong at small $\alfa$, it does
have the correct asymptotic behaviour.  Our interpretation of the RS
solution is that it should be considered as a smooth regularization of
the true mutual information, which is given by the analytical
solution, eq.~(\ref{analyticI1}), for $\alfa < \alfa_c$.  The precise
value of $\alfa_c$ cannot be determined by our techniques.  There is
numerical evidence \cite{simul98} supporting the validity of the analytical solution
and the conjecture that the $RSA$ solution is an excellent
interpolation between the small and large $\alfa$ behaviors.  The
analysis of the origin of the discrepancies between the RSA and the
analytical approaches will be the subject of a future work.

\indent
We have also explored some other schemes beyond the completely
symmetric ansatz. These are based on different types of replica
symmetry breaking ans\"atze such as the usual breaking of the symmetry
\cite{Mezard} and the separation of the first replica from the
others.  In the first case it was shown that the new solution
coincides with the symmetric one. In the second, and because of
the complexity of the problem, we have not been able to give an
explicit final result.

\section*{Note added in proof}

\indent

Part of this work was presented at the "Interdisciplinary 
Workshop on Neural Networks", W\"urzburg, Germany,(October' 95) and at 
the "Fisica Estadistica'96" meeting, Zaragoza, Spain (May'96).

\section*{Acknowledgements}

\indent

We warmly thank J-P Nadal for fruitful discussions on this paper.
This work was supported by a Spanish grant PB 96-47.
Antonio Turiel is financially supported by a FPI grant from Comunidad 
Aut\' onoma de Madrid. Elka Korutcheva was financially supported by the 
Spanish Ministry of Science and Education, and partially by Contract 608 
from the Bulgarian Scientific Foundation. E.K. also thanks the 
Department of Fundamental Physics of UNED for useful discussions during 
the preparation of the present work.

\newpage

\appendix

\vspace*{2cm}

\noindent
{\huge\bf Appendices }

\noindent
\rule[4pt]{\textwidth}{1pt}

\section{Calculation of $I_2$}
\label{I2}

\indent
To compute $I_{2}$ in the general case, we use again the 
fact that $\xiv$ and $\hv$ are deterministically related, which leads to

\be
I_{2}= \sum_{i=1}^{P} \sint{}{\Jm} \rho (\Jm ) \sint{P}{\hv } 
\rho (\hv | \Jm) \bitent{h_{i}}\;  
\ee 

\noindent
or, in terms of the Fourier transforms of the field distribution 
$\rho(\hv | \Jm)$ and of $\bitent{h_{i}}$ ($\hat{\rho} (\uv | \Jm)$ and 
$\hbitent{u_{i}}$, respectively)

\be
\label{Idos} 
I_{2}= \sum_{i=1}^{P} \sint{}{\Jm} \rho (\Jm ) \sint{P}{\uv } 
\hat{\rho} (\uv | \Jm) \hbitent{u_{i}}.
\label{compare1}  
\ee 

\noindent
The Fourier transform of the field distribution is computed in 
Appendix \ref{PgreaterN}. One has

\be 
\hr{\uv |\Jm }= e^{- 2 \pi ^{2} \uv \Deltam \uv^t }\; ,
\label{hatrho} 
\ee

\noindent
where $\Deltam = \Jm\,\Gm\,\Jm ^{t} \, \in \matreal{P}{P} $. 
After integrating over the $\Jm$ in eq.~(\ref{Idos}), we obtain:

\be I_{2}= \sum_{i=1}^{P} \sint{P}{\uv} \hr{\uv } \hbitent{u_{i}} 
\; ,\label{compare2} \ee 

\noindent
where  $\hr{\uv }$ is the characteristic function of $\rho (\hv )$. 
Although we cannot calculate the probability density of \hv, we can have 
an explicit expression for its Fourier transform by comparing 
eqs.~(\ref{compare1}) and (\ref{compare2}).
Replacing eq.(\ref{hatrho}) in eq.(\ref{compare1}) we obtain

\be \hr{\uv }= \frac{1}{\sqrt{\det{(\id{NP} + 4 \pi ^{2} \Um 
\otimes \Mm )}}}\; . \ee

\noindent
Here,

\begin{itemize}

\item
$\id{NP}$ is the identity matrix in NP dimensions .

\item
``$\bigotimes $'' stands for the tensor product between $\Um$ and \Mm.

\item
$\Mm = \Gammam\,\Gm$ is a constant, $N$ dimensional matrix.
Since $\hv $ is of order one, Tr(\Mm ) is also of order one 
(Tr stands for the trace).

\item
$\Um $ is a $P$ dimensional matrix defined as the projector 
on \uv:  
$(\Um )_{ii'} = u_{i} u_{i'} $.

\end{itemize}

\noindent
Since \hr{\uv } is invariant under similarity transformations  of \Mm,
it can be expressed as 

\be
\hr{\uv } =\frac{1}
{\prod_{j=1}^{N} \sqrt{\det{(\id{P} + 4 \pi^{2} m_{j} \Um) }}}, 
\ee 

\noindent
where $\{ m_{j} \}, \,\, j=1,...,N$,  is the set of eigenvalues 
of \Mm. The matrix $\Um $ has  only one non-zero eigenvalue, which is 
$\modd{\uv } = \uv \cdot \uv$ and thus:

\be
\label{l4} 
\hr{\uv } = e^{-\frac{1}{2} \ln{\prod_{j=1}^{N} (1+4 \pi ^{2} m_{j} 
\modd{\uv } )}}\; . \ee 

\indent
The computation of $I_{2}$ does not need the whole joint distribution 
$\rho (\hv )$  but only the marginals $\rho (h_{i}); i = 1, \ldots, P$. 
By permutation symmetry, it is obvious that all of them are given by the 
same function. Let us compute for example  $\rho (h_{1})$. Its Fourier 
transform is
 
\be 
\hr{u_{1}}=\hr{u_{1},0,\, ^{P-1\prime}_{\,.\,.\,.},0}
 \, = \, e^{- \frac{1}{2} \sum_{j=1}^{N} \ln{(1+4 \pi ^{2} m_{j} 
(u_{1})^{2}) } }\;  
\ee

\noindent
and since all the marginals are the same function, all the terms in 
$I_{2}$ are the same. Then $I_{2}$ reads

\be 
I_{2} = P \int_{-\infty}^{\infty} dh \;\rho (h) \bitent{h} \; . 
\ee  

\indent
So far there is no hypothesis upon the matrix $\Mm$. Particularly 
interesting is the case in which all the $m_{j}$'s are 
of the same order, namely, of order $1/N$ (as we have already said,  
Tr(\Mm) is O(1)~). In this particular case,

\be 
\hr{u} \sapprox{N\gg 1}  e^{ - \frac{1}{2} \sum_{j=1}^{N} 4 \pi 
^{2} m_{j}  u^{2} } + O(e^{-N}) = e^{- 2 \pi ^{2} \sbM u^{2} } + O(e^{-N}) , 
\ee 

\noindent
where $\bM = $Tr(\Mm). In the thermodynamic limit the term $O(e^{-N})$
becomes negligible and $\rho(h)$ is~:
	
\be\lim_{N \rightarrow \infty } \rho (h) = \frac{e^{- h^{2} /(2 
\sbM )}} {\sqrt{2 \pi \bM }}.
\ee 

\noindent
(Note that this expression makes explicit the reason why  $\bM = O(1)$). 
The conditional output entropy now is~:

\be I_{2}= P \int_{- \infty}^{\infty} dz\; \frac{e^{-z^{2}}}{\sqrt{\pi}} 
\;\bitent{\sqrt{2 \bM }\, z} \; .
\label{generalI2}
\ee

\section{ Computation of \hr{\uv |\Jm } for $P > N$}
\label{PgreaterN}

\indent

We define the Fourier transform of a function $F(\hv )$ 
as the function $\hat{F} (\uv )$ given by~:

\be 
\hat{F} (\uv ) = \sint{P}{\hv} F(\hv ) e^{-2 \pi i \hv 
\cdot \uv}\; . 
\label{Ftdef} 
\ee

\indent
The evaluation of $\hr{\uv |\Jm }$ in the case $P\leq N$ is 
simple. This is because for almost every $\Jm$ the random vector $\hv$ 
follows a Gaussian distribution with correlation matrix $\Deltam = 
\Jm\:\Gm\Jm^{t}$ and det(\Deltam)$\neq 0$. Then,

\be \hr{\uv|\Jm } = e^{-2\pi^{2}\uv \Deltam\uv^t} . \ee

\indent
We now prove that this equation is still true when $P>N$. Let us first 
notice that in this case det(\Deltam) is necessarily null, 
and consequently the random vector $\hv$ is not Gaussian.

\indent
Let us compute $\prob(\vv|\Jm) $ for the particular vector
$\vv=(1,1,...,1)$. Denoting this as $\prob$ we have:

\be \prob = \sint{N}{\xiv } \frac{e^{-\frac{1}{2} \xiv (\Gm )^{-1} \xiv^t 
}}{\sqrt{\det{(2 \pi \Gm )}}} \prod_{i=1}^{P} f(h_{i})\;\;  
= \;\; \sint{P}{\uv} \hat{\rho}(\uv |\Jm ) \prod_{i=1}^{P} \hat{f}(u_{i})
\label{appA1}
\ee

\noindent
with $h_{i} = \sum_{j=1}^{N} J_{ij}\:\xi _{j}$ and $u_i$ being its
conjugate Fourier variable. For $P > N$, the first N components of 
$\hv$ are independent random variables and the other $P-N$ depend upon
the former (for almost every $\Jm $).

\indent
We split the matrix $\Jm$ into two matrices: $\Km\in\matreal{N}{N}$, 
$K_{jj^{\prime }}= J_{jj^{\prime} }$; and $\Lm\in\matreal{P-N}{N} $, 
$L_{kj}=J_{N+k,j}, k=1,...,P-N;\: j=1,...,N $:

\[
\Jm\; =\; \left(\begin{array}{c} \Km \\ 
\\
\Lm 
\end{array}
\right)
\hspace*{-33pt}\rule[3pt]{25pt}{0.5pt} 
\]

\noindent
and for almost every $\Jm $ , $\Km $ is invertible. 
Then, we split $\hv =(\hvu ,\hvd )$, $\hvu\in\Real^{N} $ and 
$\hvd\in\Real^{P-N}$. Moreover, $\hvu $ is gaussianly distributed with 
zero mean and covariance matrix $\Deltam_{0}=\Km\:\Gm\:\Km^{t}$, and 
$\hvd = (\Lm\:\Km^{-1})\;\hvu $.

\indent
In this way, we obtain $\prod_{i=1}^{P} f(h_{i}) = \prod_{j=1}^{N} 
f(h_{j}^{0} )\;\prod_{k=1}^{P-N} f([(\Lm\:\Km^{-1})\;\hvu ]_{k}) $, and
hence $\prob $ can be written as:

\be \prob = \sint{N}{\hvu } \frac{e^{-\frac{1}{2} 
\hvu (\Deltam_{0})^{-1}\hvu^t }}{\sqrt{\det{(2\pi\Deltam_{0})}}} 
\prod_{j=1}^{N} f(h_{j}^{0} )\;\prod_{k=1}^{P-N} f([(\Lm\:\Km^{-1})\;\hvu 
]_{k} ) \; .\ee

\indent
If $g(\hv )$ is a function of vectorial argument, and $f(x)$ has real 
argument, we have that:

\be ( g(\hv ) \, f(\vec{a} \cdot \hv ) )^{\wedge } (\uv ) = 
\int_{-\infty }^{\infty } dc\; \hat{g} (\uv -c\, \vec{a} ) \hat{f} (c)\; 
, \ee

\noindent
where the hat symbol stands for the Fourier transform and $\vec{a} $ is an 
arbitrary constant vector. It should be noted that $\hat{g}$ is a 
multidimensional Fourier transform while $\hat{f}$ is the 
one-dimensional Fourier transform.

\indent
Let us denote by $\vec{d}_{k}$ the $(P-N)$ N-dimensional vectors defined by
the rows of $\Lm\:\Km^{-1}$. Applying the previous formula to the expression 
for $\prob $, and after using the Bessel-Plancherel identity, we obtain:

\be \prob = \sint{N}{ \uv^{\,\prime} } e^{-2 \pi^{2} \uv^{\,\prime} 
\Deltam_{0} 
\uv^{\,\prime t} }  \sint{P-N}{\vec{c}}\prod_{j=1}^{N} \hat{f} 
(u_{j}^{\prime} - 
\sum_{k=1}^{P-N} c_{k} (\vec{d}_{k} )_{j} )\;\prod_{k=1}^{P-N} 
\hat{f}(c_{k}) \; .\ee

\noindent
Interchanging now the order of integrations and performing the change of
variables $\uv^{\, 0}$, related via $\uv^{\,\prime } = \uv^{\, 0} +
\sum_{k=1}^{P-N} c_{k} \vec{d}_{k} $, we obtain:

\ba \prob =\sint{P-N}{\vec{c}} \sint{N}{\uv^{\, 0}}
\prod_{j=1}^{N} \hat{f}(u_{j}^{0}) \:\prod_{k=1}^{P-N} \hat{f}(c_{k}) 
\nonumber
\\
 e^{-2 \pi^{2} (
\uv^{\, 0} \Deltam_{0} \uv^{\, 0 t} + \sum_{k,k^{\prime}=1}^{P-N} c_{k} 
c_{k^{\prime}} \vec{d}_{k} \Deltam_{0} \vec{d}_{k^{\prime}}^t + 2 
\sum_{k=1}^{P-N} c_{k} \uv^{\, 0} \Deltam_{0} \vec{d}_{k}^t )}
. 
\label{appA2}
\ea

\indent
It is convenient to combine $\uv^{\, 0} $ and $\vec{c}$ in a single P 
dimensional vector $\uv =(\uv^{\, 0},\vec{c}) $. Expressing
eq.~(\ref{appA2}) in terms of this vector, we can use the
vectors $\vec{d}_{k} $ to simplify the bilinear expression in the 
exponent as:

\be \prob = \sint{P}{\uv } e^{-2\pi^{2}\uv \Deltam\uv^t } \prod_{i=1}^{P} 
\hat{f} (u_{i} )\;\; , \;\;\Deltam = \Jm\:\Gm\Jm^{t}\; , \ee

\noindent
that depends only on $\Deltam$. From the right hand side of
eq.~(\ref{appA1}) we have

\be \hr{\uv|\Jm } = e^{-2\pi^{2}\uv\Deltam\uv^t} \; .\ee

\section{RSA derivation for $I_{1}$ }
\label{RSAapp}

\indent

We now derive the Saddle Point (SP) equations. These are then
simplified by using the Replica Symmetry Ansatz (RSA)
\cite{Mezard}. The first order parameters are the overlap of two
replicated Fourier transforms of the local field:

\be \tilde{U}_{\ai\bi} = \frac{1}{P} \uv_{\ai }\cdot \uv_{\bi }\; ,\: 
\ai ,\bi =0,...,n  \label{overlap} \ee

\noindent
The pre-factor is taken in order to ensure that it is of order one. These 
are the elements of a matrix $\Utm \in \matreal{(n+1)}{(n+1)}$ . Then, the
Fourier transform of the joint distribution of the replicated local fields, 
eq.~(\ref{hatrho_1}), can be expressed in terms of this matrix as:

\be
\label{l5} 
\hr{\{\uv ^{\ai} \}} = \prod _{\ai\leq\bi} \int_{- \infty 
}^{\infty } dU_{\ai\bi} \delta (U_{\ai\bi} - \frac{1}{P} 
\uv_{\ai} \uv_{\bi} ) e^{-\frac{1}{2} \sum_{j=1}^{N} 
\ln{\det{[\:\id{(n+1)} + 4 \pi ^{2} P m_{j} \Utm\: ] }}} \; .\ee 

\noindent
Now we introduce an order parameter $\Vtm$, conjugated to $\Utm$.  To
linearize the quadratic form in the $\uv^{\ai}$'s we will use $P$ new
variables $\wv_i$, which are $(n+1)$-dimensional vectors.

\indent
After substituting eq.~(\ref{l5}) in eq.~(\ref{momentprob}), we can 
perform the integrals over the $\uv _{\alpha }$. 
Since these integrals are the 
anti-Fourier transforms of the $\hat{f}(u_i^{\ai})$, the $\ll \prob 
^{n+1} \gg$ can be expressed in terms of the product of the transfer 
functions simply~:

\ba \ll \prob ^{n+1} \gg = \int \prod_{\ai\leq\bi} ( 
- i P  d \tilde{V}^{\ai\bi}\: d \tilde{U}_{\ai\bi} )
\nonumber \\
 e^{ 2 \pi P \sum_{\ai\leq\bi} \tilde{U}_{\ai\bi}
\tilde{V}^{\ai\bi} 
-\frac{1}{2} \sum_{j=1}^{N} \ln{\det{[\:\id{(n+1)} + 4 \pi ^{2} m_{j} 
\Utm\: ] }} }
 \nonumber \\
\prod_{l=1}^{P} \int d^{n+1} \wv_{l}\;\; \frac{e^{-\frac{1}{2} \wv_l 
(\Vtm )^{-1} \wv_l^t }}{\sqrt{\det{(2 \pi \Vtm )}}} \prod_{\ai=0}^{n}
f(\frac{w_{l}^{\ai}}{\sqrt{\pi }} )\; .\ea 

\noindent
($i\equiv\sqrt-1$). This can be written as:

\ba 
\label{l3}
\ll  \prob ^{n+1} \gg  = \int \prod_{\ai\leq\bi} 
(-i P d \tilde{V}^{\ai\bi}\: d \tilde{U}_{\ai\bi})
\nonumber \\
e^{ 2 \pi P \sum_{\ai\leq\bi} 
\tilde{U}_{\ai\bi} \tilde{V}^{\ai\bi} -\frac{1}{2}
\sum_{j=1}^{N} \ln{\det{[\:
\id{(n+1)} + 4 \pi ^{2} P m_{j} \Utm\: ]}} + P \ln{ Z( \Vtm ) } } 
\; ,\ea
 
\noindent
where

\be Z(\Vtm ) = \int d^{n+1} \wv\;\;  \frac{e^{-\frac{1}{2} \wv (\Vtm )^{-1} 
\wv^t }}
{\sqrt{\det{(2 \pi \Vtm )}}} \prod_{\ai= 0}^{n} f(\frac{w^{\ai}}{\sqrt{\pi }}) .
\label{Zdef}
\ee

\noindent
In the large $N$ limit ($\alfa =\frac{P}{N}$ fixed), the
integrals over $\Utm$ and $\Vtm$ in eq.~(\ref{l3}) can be solved by the
SP method. This gives:

\be \ll \prob ^{n+1} \gg \approx e^{G( \Utm _{0}, \Vtm _{0} )} 
\label{RSAmoments} 
\ee

\noindent
where $\Utm_0$ and $\Vtm_0$ are the SP values and

\be G = 2 \pi P \sum_{\ai\leq\bi} \tilde{U}_{\ai\bi} 
\tilde{V}^{\ai\bi} 
-\frac{1}{2} \sum_{j=1}^{N} \ln{\det{[\id{(n+1)} +4 \pi ^{2} P m_{j} 
\Utm ]}} + P \ln{Z(\Vtm )} \label{RSAG} .\ee

\noindent
The RSA is:

\be \left\{ \begin{array}{cccc}
\tilde{U}_{\ai\ai}^{0} & = & U_{0}, & \forall\ai \\
\tilde{U}_{\ai\bi}^{0} & = & U_{1}, & \forall\ai\neq\bi 
\end{array} 
\right.
\hspace*{1cm} \mbox{and} \hspace*{1cm}
\left\{ \begin{array}{cccc}
\tilde{V}_{\ai\ai}^{0} & = & V_{0}, & \forall\ai\\
\tilde{V}_{\ai\bi}^{0} & = & V_{1}, & \forall\ai\neq\bi 
\end{array} 
\right. \label{RSAparam} \ee

\indent
The starting point is eq.~(\ref{RSAmoments}), where the function $G$
is evaluated with the RSA given in eq.~(\ref{RSAparam}). Defining the
matrix $\unom\in\matreal{(n+1)}{(n+1)}$ as $ (\unom )_{\ai\bi} = 1
\;\forall\;\ai ,\bi $, $\Vtm$ can be expressed as:

\be \Vtm = (v_{0} -\frac{1}{2} v_{1})\id{(n+1)} +\frac{1}{2} v_{1} \unom \ee

\noindent
and its inverse is:

\be (\Vtm )^{-1} = x_{0} \id{(n+1)} + x_{1} \unom , \ee

\noindent
where $x_{0}= \frac{1}{v_{0}-\frac{1}{2} v_{1}} $ and $x_{1}= 
-\frac{\frac{1}{2} v_{1}} {(v_{0}-\frac{1}{2} v_{1} )( v_{0} + 
\frac{n}{2} v_{1} ) }$. This form of $(\Vtm )^{-1}$ allows us to express
$Z(\Vtm )$ (eq.~(\ref{Zdef})) in a more convenient way:

\ba Z(\Vtm )= (2\pi)^{-\frac{n+1}{2}}\, 
x_{0}^{\frac{n}{2}}\,\sqrt{x_{0}+(n+1) x_{1} }
\nonumber
\\
\int d^{n+1} \wv \;\;
e^{-\frac{1}{2} x_{0} \sum_{\ai =0}^{n} w_{\ai}^{2} -\frac{1}{2} x_{1}
(\sum_{\ai =0}^{n} w_{\ai})^{2}} \; \prod_{\ai =0}^{n} 
f(w_{i}^{\ai }/\sqrt{\pi }) \; .\ea

\indent
Notice that  $Z(\Vtm )\rightarrow \frac{1}{2}$ as $n\rightarrow 0$. We
now expand $Z(\Vtm )$ up to order $n$, $Z\approx
\frac{1}{2} +n\,h(x_{0},x_{1} )$, $h = \partial Z/\partial n |_{n=0}$.
Up to this order we obtain:

\ba G \approx 2\pi P (n+1) v_{0} u_{0} + \pi P\: n\: u_{1} v_{1} 
-\frac{1}{2} \sum_{j=1}^{N} (1 +4 \pi^{2} P m_{j} u_{0}) -\frac{n}{2} 
\sum_{j=1}^{N} 
\frac{ 4 \pi^{2} P m_{j} u_{1}}{1+4\pi^{2} P m_{j} u_{0} }
\nonumber
\\
-\frac{n}{2} \sum_{j=1}{N}\ln{(1+4\pi^{2} P m_{j} (u_{0}-u_{1} ))} 
-P \ln{2} + 2 P n h(x_{0}, x_{1} ) \; . 
\label{appB1}
\ea

\indent
The SP equations extremize G with respect to its 
variables. From the SP equation $\partial G/\partial v_{0} = 0$ one
obtains that $u_0$ is linear in $n$: $ u_{0} = n \tilde{u} $. Replacing
eq.~(\ref{appB1}) in eq.~(\ref{RSAmoments}), and this in
eq.~(\ref{limitI1}) we have:

\be I_{1}^{RSA} = -2 \pi P v_{0} \tilde{u} +\pi P v_{1} u + 2\pi^{2} P \bM 
\tilde{u} -2 \pi^{2} P \bM u +\frac{1}{2} \sum_{j=1}^{N} \ln{(1+4\pi^{2} 
P m_{j} u )} - 2 P h(x_{0}, x_{1})  \;  
\label{I1previo}
\ee

\noindent
where $u= -u_{1} $. The SP equations are :

\be \left\{ \begin{array}{ccc}
v_{0} & = & \pi\bM \\ \\
v_{1} & = & 2\pi\bM - 2\pi \sum_{j=1}^{N} \frac{m_{j}}{1+4\pi^{2} P m_{j} 
u}\\ \\
\tilde{u} & = & -\frac{1}{\pi}\partial h/\partial v_{0} \\ \\
u & = & \frac{2}{\pi} \partial h/\partial v_{1} 
\end{array} \right. \;  \ee

\noindent
$\tilde{u} $ is a Lagrange multiplier, that can be 
easily removed by substituting the value of $v_0$ in $I_{1}^{RSA}$.

\indent
The evaluation of $h(x_{0},x_{1})$ requires some care. We express $Z$ as:

\be Z(\Vtm )= (2\pi)^{-\frac{n+1}{2}}\, 
x_{0}^{\frac{n}{2}}\,\sqrt{x_{0}+(n+1) x_{1} }\;\int_{-\infty}^{\infty} dx
\frac{e^{-\frac{1}{2} x^{2}}}{\sqrt{2\pi}} [ {\cal L \/} (x) ]^{n+1} \; , \ee

\noindent
where

\be {\cal L \/} (x) = \int_{-\infty}^{\infty} dw\; e^{-\frac{1}{2} x_{0} 
w^{2} +\: i \sqrt{x_{1}} \: x w } f(x/\sqrt{\pi} ) \; . \ee

\noindent
Computing the term of order $n$ of $Z$ we obtain:

\be h(x_{0},x_{1}) = -\frac{1}{2} \ln{2} -\frac{1}{2} 
\sqrt{\frac{x_{0}+x_{1}}{2\pi x_{0}}} \tilde{M} (x_{0}, x_{1})\;  
\label{hx0x1}
\ee

\noindent
with

\be \tilde{M}(x_{0},x_{1}) = \int_{-\infty}^{\infty} dz\;
e^{-\frac{x_{0}+x_{1}}{2 x_{0} } z^{2}} \tilde{F}(\tilde{g} (z) )\; ,  \ee

\noindent
where the function $\tilde{F}(y)$ is defined by: 

\be \tilde{F}(y) = \frac{1}{2} 
\ln{\frac{1-y}{1+y}} -\frac{1}{2} \ln{(1-y^{2} )} \ee

\noindent
and its argument is~:

\be \tilde{g} (z) = \sqrt{\frac{2 x_{0}}{\pi }} \, 
e^{\frac{1}{2}\frac{x_{1}}{x_{0}} z^{2} } \int_{-\infty}^{\infty} dw\; 
e^{-\frac{1}{2} x_{0} w^{2} }\, \sinh{(\sqrt{-x_{1}}\: z w )} 
f(w/\sqrt{\pi } )\; . \ee

\noindent
Substituting $h(x_0,x_1)$ given above in eq.~(\ref{I1previo}), we obtain:

\be I_{1}^{RSA}= P \ln{2} + P \sqrt{\frac{x_{0}+x_{1}}{2 \pi x_{0}}} \, 
\tilde{M}
(x_{0},x_{1}) +\frac{1}{2} \sum_{j=1}^{N} \ln{(1 + 4\pi^{2} P m_{j} u)} 
-\frac{1}{2} \sum_{j=1}^{N} \frac{4 \pi^{2} P m_{j} u}{1+4 \pi^{2} P 
m_{j} u} \; . 
\label{appB2}
\ee

\noindent
and the SP equations become:

\be \left\{ \begin{array}{ccc}
x_{0} & = & 1/(\pi \sum_{j=1}^{N} \frac{m_{j}}{1+4\pi^{2} P m_{j} u} )\\ \\
x_{1} & = & \frac{1}{\pi \bM } - x_{0} \\ \\
u & = & -\frac{x_{0}^{2}}{(2\pi )^{3/2}} \: (\frac{\partial}{\partial x_{0}}
-\frac{\partial }{\partial x_{1}} ) [ \sqrt{\frac{x_{0} +x_{1}}{x_{0}}} 
\tilde{M} (x_{0},x_{1} ) ]
\end{array} \right. 
\label{appB3}
\ee

\indent
We can substitute in $I_1^{RSA}$ one of the parameters, for example
$x_{0}$ . Defining $x =- \pi \bM\: x_{1}$, $s= 4 \pi^{2} \bM\:\alfa u $ 
and rearranging conveniently eqs.~(\ref{appB2}) and (\ref{appB3}), we 
finally obtain eqs.~(\ref{RSAI1}) and (\ref{SPeqs}).

\section{Analytical derivation of $I_{1}$ }
\label{Analyticapp}

\indent

This exact calculation starts from eq.~(\ref{startmethod2}). We assume 
that all the eigenvalues $\{ m_{j} \}_{j=1,...,N}$ of the matrix $\Mm$ 
are at most of order $\frac{1}{N}$. We define now $\Utm $ in a slightly 
different way from  eq.~(\ref{overlap}):

\be \tilde{U}_{\ai\bi } = \uv^{\ai} \cdot \uv ^{\bi}. \ee

\noindent
These elements are of order $P$ and hence $m_{j}\Utm$ is of order $\alfa$.
$\hr{\{\uv^{\ai}\}}$ is computed as in subsection \ref{I1RSA} . Then,
the logarithm in the exponent of eq.~(\ref{l5}) can be expanded around the 
identity matrix (what can be donde if $\alfa $ is less than
$\frac{1}{4 \pi ^{2} \bM}$ times a geometrical factor of order one,
that depends on $\Ggm $)

\be \hr{\{\uv^{\ai}\}} = \prod_{m=1}^{\infty} e^{\frac{1}{2} (-1)^{m}
\frac{(4 \pi ^{2})^{m}}{m} Tr(\Mm^{m}) Tr(\Utm^{m})} \; . \ee 

\noindent
Let us remark that {\it this is not an approximation}. It 
is an {\it exact \/} derivation valid in a (undetermined) range of
values $\alfa $ {\it of order one \/}. We can 
alternatively write this in the following form~:

\be \hr{\{\uv^{\ai}\}} = e^{ - 2\pi^{2}\sbM\sum_{\ai =0}^{n} 
\modd{\uv^{\ai}} } \:\prod_{m=2}^{\infty} e^{\frac{1}{2 m} (-4 
\pi^{2} \sbM)^{m} N Tr(\Ggm^{m}) Tr((\Utm /N)^{m}) } \; . \ee 

\noindent
The second factor can now be expanded, leading to polynomials in
traces of powers of $\Utm /N$.

\indent
Given a function $F(\{\uv^{\ai}\})$, we now define its average with
the transfer function (or shortly, its transfer-average) as:

\be \ta{F} = \int \prod_{\ai = 0}^{n} d^{P}\!\uv^{\ai} \: 
e^{-2 \pi^{2}\sbM \modd{\uv^{\ai}} } F(\{ \uv^{\ai}\} )
\prod_{\ai =0}^{n} \prod_{i=1}^{P} \hat{f} (u_i^{\ai} ) \; . \ee

\noindent
Notice that

\be \ta{1} = 2^{-P (n+1)} .\ee

\noindent
This property comes from the fact that $f(x)+f(-x)=1$, and so $f(x)=
1/2 + a(x)$, where $a(x)$ is an odd function bounded by $-\frac{1}{2}$
and $\frac{1}{2}$.  Notice that, although this will not be required
here, in the physically reasonable cases $a(x)$ is an increasing and
almost everywhere continuous function, $0\leq a(x) \leq \frac{1}{2},\;
x > 0$ such that $a(x)\rightarrow \frac{1}{2} $ when
$x\rightarrow\infty$.

\indent
In terms of these transfer-averages, the moments read:

\be \ll \prob^{n+1} \gg = 2^{-P (n+1)} + \sum_{\sigma =2}^{\infty} \sum_{ 
_{_{_{_{_{_{_{s_{1} t_{1}+...+s_{\lambda } t_{\lambda } = \sigma }}}}}}}
\gaga\gaga  2\leq s_{1} < ... < 
s_{\lambda } } C_{s_{1},...,s_{\lambda }}^{t_{1},...,t_{\lambda }}
\:\Lambda _{s_{1},...,s_{\lambda }}^{t_{1},...,t_{\lambda }} \; ,
\label{moment_ta}
\ee 

\noindent
where

\be C_{s_{1},...,s_{\lambda }}^{t_{1},...,t_{\lambda }} = \frac{(-4 \pi 
^{2})^{\sigma }}{t_{1}!\, ...t_{\lambda }!} 
(\frac{N \bM}{2 })^{^{^{^{t_{1}+...+t_{\lambda} }}}}\: \frac{(Tr[(\Ggm 
)^{s_{1}}])^{t_{1}}}{s_{1}^{t_{1}}}\: ... \:\frac{(Tr[(\Ggm )^{s_{\lambda 
}}])^{t_{\lambda }}}{s_{\lambda }^{t_{\lambda }}} \ee

\noindent
and

\be \Lambda_{s_{1},...,s_{\lambda }}^{t_{1},...,t_{\lambda }} =
\ta{ (Tr[ (\Utm /N)^{s_{1}} ])^{t_{1}} \: ... \: (Tr[ (\Utm 
/N)^{s_{\lambda }} ])^{t_{\lambda }} } \; .\ee

\noindent 
These transfer-averages have a very simple expression in the 
thermodynamic limit, what allows us to rearrange the whole expression 
in a convenient way. First, we must notice that:

\be \int_{-\infty}^{\infty} du\; e^{-2\pi^{2}\sbM u^{2}}\: u^{2 r} 
\hat{f}(u) = \frac{1}{2} \delta_{0r} 
\label{appC1}
\ee

\noindent
because $f(x)=\frac{1}{2}+a(x)$ and $a(x)$ is odd. We now prove a
factorization property , eq.~(\ref{factorization}), of the
transfer-averages of traces of $\Utm$ that will be useful to
computate of $\ll\prob^{n+1}\gg$.

\indent
The trace of the r-th power of $\Utm $ can be written as:

\be Tr(\Utm^r )= \sum_{\ai_{1},...,\ai_{r}\, = 0}^{n}
\sum_{i_{1},...,i_{r} = 1}^{P} 
u_{i_{1}}^{\ai_{1}}\, u_{i_{2}}^{\ai_{1}}\, u_{i_{2}}^{\ai_{2}}\,
u_{i_{3}}^{\ai_{2}}\, u_{i_{3}}^{\ai_{3}}\, u_{i_{4}}^{\ai_{3}}\;
... \; u_{i_{r-1}}^{\ai_{r-1}}\, u_{i_{r}}^{\ai_{r-1}}\, 
u_{i_{r}}^{\ai_{r}}\, u_{i_{1}}^{\ai_{r}} .
\label{summation}
\ee

\noindent
After taking the transfer-average on this expression, one should notice that 
the contribution of each term does not depend on the particular indeces 
$i$'s and $\alfa$'s present in that term. It only depends on the number
of different variables in the term and the power of each variable.
This is due to the independency and permutation symmetry of the
$u$'s. It is possible to rearrange eq.~(\ref{summation}), 
expressing it as the sum of each different contribution times a
combinatorial factor. This factor is the number of terms giving that
particular contribution. Since the contributions themselves are of order
one, the thermodynamical limit is determined by the combinatorial factors.
In this limit, $P\rightarrow\infty$ ($r$ kept finite) and the
combinatorial factors scale as $P$ raised to the number of
non-repeated indeces $i$. Then, no more than
two $u$'s can be equal. Defining

\be \lambda (\bM )= \int_{-\infty}^{\infty} du\; e^{-2\pi^{2}\sbM u^{2}} 
\; u \hat{f} (u) \ee

\noindent
and considering eq.~(\ref{appC1}) for $r=0$ and $r=1$, the 
transfer-average of eq.~(\ref{summation}) can be expressed in terms of 
$\lambda$:

\be \ta{Tr[(\Utm /N)^{r}]} =_{_{_{_{\ga N\rightarrow\infty\; }}}} 
2^{-P (n+1)} \chi_{r} \; , \ee 

\noindent
where

\be \chi_{r} = ( n^{r} +(-1)^{r} n ) (2 \lambda )^{2 r} \alfa^{r} \; .
\label{chidef}
\ee 

\indent
By similar arguments, one can prove a useful factorization property.
For the product of two traces we have:

\be \ta{Tr[(\Utm /N)^{r}]\: Tr[(\Utm /N)^{s}] } = 2^{-P (n+1)}\:\chi_{r}\: 
\chi_{s} \; 
\label{factorization}
\ee 

\noindent
and  a similar factorization holds for the product of an arbitrary number
of traces. Recalling eq.~(\ref{moment_ta}), this property allows us to
write:

\be \ll\prob^{n+1}\gg = 2^{- P (n+1)}\; \exp\left[N \sum_{m=1}^{\infty}
\frac{1}{2m} (-4 \pi^{2} \sbM )^{m} Tr[(\Ggm )^{m}]  \:\chi_{m} 
\right]
 \; .\ee

\noindent
Substituting the explicit values of the $\chi $ 's, eq.~(\ref{chidef}), and
preforming the sum, we have:

\be \ll\prob^{n+1}\gg = 2^{-P (n+1) } \; e^{-\frac{1}{2} Tr[\:\ln{(\id{N} 
+16 
\pi^{2} \lambda^{2} n P \Mm )}\: ] } \, e^{-\frac{n}{2} Tr[\:\ln{(\id{N} 
- 16 \pi^{2}\lambda^{2} P \Mm )}\: ] } \; .
\label{appC2}
\ee 

\noindent
These moments can be expressed in a more useful way. Defining $k$ by:

\be  k = \int_{-\infty}^{\infty} dy\;\; y\, e^{-\frac{y^{2}}{2}} 
f(\sqrt{\bM}\; y) \; ,\ee 

\noindent
we have $\lambda^{2} = -\frac{k^{2}}{8 \pi^{3} \sbM   }$. Using this
relation in eq.~(\ref{appC2}), we finally obtain eq.~(\ref{analyticI1}).

\newpage

\listoffigures

\end{document}